\documentclass{article}

\usepackage[preprint]{neurips_2020}

\usepackage[utf8]{inputenc}
\usepackage[T1]{fontenc}
\usepackage{hyperref}
\usepackage{url}
\usepackage{booktabs}
\usepackage{array}
\usepackage{amsfonts}
\usepackage{nicefrac}
\usepackage{microtype}
\usepackage{import}
\usepackage{amsmath}
\usepackage{graphicx}
\usepackage{bbm}

\usepackage{caption}

\newcommand{\Lce}{\mathcal{L}_\textnormal{cross-entropy}}
\newcommand{\LAUC}{\mathcal{L}_\textnormal{AUC}}
\newcommand{\LAUCprev}{\mathcal{L}_\textnormal{AUC-prev}}
\newcommand{\LleftAUC}{\mathcal{L}_\textnormal{leftAUC}}
\newcommand{\LlogAUC}{\mathcal{L}_\textnormal{logAUC}}
\newcommand{\Xcurr}{\mathcal{X}_\textnormal{curr}}
\newcommand{\Ycurr}{\mathcal{Y}_\textnormal{curr}}
\newcommand{\Xprev}{\mathcal{X}_\textnormal{prev}}
\newcommand{\Yprev}{\mathcal{Y}_\textnormal{prev}}

\renewcommand{\t}[1]{\textnormal{#1}}

\usepackage{color}

\renewcommand{\bf}{\bfseries}

\definecolor{otherloss}{rgb}{0.7,0.7,0.7}
\newcolumntype{g}{>{\color{otherloss}}l}

\newcommand{\ifelse}[3]{
\begingroup
\begin{cases}
  {#1}, & \text{if } {#2} \\
  {#3}, & \text{otherwise}
\end{cases}
\endgroup
}

\title{Deep Learning for Virtual Screening: \\ Five Reasons to Use ROC Cost Functions}

\author{
  Vladimir Golkov\,$^{\text{1}}$,
  Alexander Becker\,$^{\text{1}}$,
  Daniel T. Plop\,$^{\text{1}}$,
  Daniel \v{C}uturilo\,$^{\text{1}}$,
  Neda Davoudi\,$^{\text{1}}$,\\ 
  \textbf{
    Jeffrey Mendenhall\,$^{\text{2}}$,
    Rocco Moretti\,$^{\text{2}}$,
    Jens Meiler\,$^{\text{2,3}}$,
    Daniel Cremers\,$^{\text{1}}$
  }
  \\
  $^1$ Computer Vision Group, Technical University of Munich, Germany\\
  $^2$ Center for Structural Biology, Vanderbilt University, USA\\
  $^3$ Institute for Drug Discovery, Leipzig University, Germany\\
  \footnotesize{\texttt{ \{vladimir.golkov, alexander.becker, daniel.plop, daniel.cuturilo, neda.davoudi, }}\\
  \footnotesize{\texttt{ cremers\}@tum.de, \{jeffrey.l.mendenhall, jens.meiler\}@vanderbilt.edu,}}\\
  \footnotesize{\texttt{ rmorettiase@gmail.com }}
}

\begin{document}

\maketitle

\begin{abstract}
Computer-aided drug discovery is an essential component of modern drug development. Therein, deep learning has become an important tool for rapid screening of billions of molecules \emph{in silico} for potential hits containing desired chemical features. Despite its importance, substantial challenges persist in training these models, such as  
severe class imbalance, high decision thresholds, and lack of ground truth labels in some datasets. In this work we argue in favor of directly optimizing the receiver operating characteristic (ROC) in such cases, due to its robustness to class imbalance, its ability to compromise over different decision thresholds, certain freedom to influence the relative weights in this compromise, fidelity to typical benchmarking measures, and equivalence to positive/unlabeled learning. We also propose new training schemes (coherent mini-batch arrangement, and usage of out-of-batch samples) for cost functions based on the ROC, as well as a cost function based on the logAUC metric that facilitates early enrichment
(i.e.~improves performance at high decision thresholds, as often desired when synthesizing predicted hit compounds).
We demonstrate that these approaches outperform standard deep learning approaches on a series of PubChem high-throughput screening datasets that represent realistic and diverse drug discovery campaigns on major drug target families.
\end{abstract}

\section{Introduction}
\label{introduction}

Drug discovery is a long, complex, and expensive process, through which new chemical compounds can be identified and that, if successful, may lead to new pharmaceutical drugs. 
In the testing process of compounds to be identified as potential drugs, we generally have many more negative samples than positive ones.
A preselection \emph{in silico} (virtual screening) using machine learning must select only a small percentage among millions of candidates, because running subsequent \emph{in vitro} experiments on each preselected compound is expensive. In other words, machine learning for virtual screening is used with high decision thresholds.
A particularly appropriate family of cost functions for this task is based on the receiver operating characteristics (ROC) curve. These functions are inherently robust to class imbalance and some of them are specifically optimized for high decision thresholds.
The ROC curve for a binary classification problem plots the true positive rate (TPR) as a function of the false positive rate (FPR). Each point on the curve is obtained by choosing a classification threshold.
The area under the ROC curve (AUC) is a measure of classifier performance.

\subsection{Reasons to use ROC-based cost functions in drug discovery}
\label{sec:reasons}

In the following we outline five important reasons to use ROC-based cost functions (i.e.~to optimize the AUC statistic or similar statistics relatively directly) in virtual screening, instead of using the typical cross-entropy cost function.

\paragraph{Class imbalance}
\label{subsub:imbalance}
Typical cost functions such as cross-entropy do not work well when applied to datasets where class sizes are strongly imbalanced and classes are not easily separable. The minority class has little contribution to the cross-entropy cost function, making it inexpensive for the classifier to 
misclassify most of the minority class.
In contrast, AUC-based cost functions sum over positive-negative sample pairs (rather than over individual samples), resulting in every class appearing equally often in the cost term. This makes AUC-based cost functions immune to class imbalance.

\paragraph{High decision thresholds}
Due to the high cost of \emph{in vitro} experiments, only a small percentage of samples can be selected, i.e.~the decision threshold must be high.
The left part of the ROC corresponds to such high decision thresholds.
Hence, \emph{optimizing} this part of the ROC specifically targets our goals.
This is why advanced quality metrics for virtual screening focus on the left part of the ROC \citep{mysinger2010rapid}.
In Section~\ref{sec:logauc}, we introduce a novel ROC-based cost function that focuses on maximizing the area under the left part of the ROC curve.

\paragraph{Not knowing the exact threshold in advance}
In virtual screening we may not know the decision threshold for classification in advance. If this is the case, it is reasonable to consider all realistic thresholds when measuring or optimizing classifier performance. The ROC (and AUC) consolidate classifier quality over all possible thresholds.
The left part of the ROC (and quality metrics that focus on it) consolidate classifier quality across various high decision thresholds.

\paragraph{Better benchmarking results}
Evaluation metrics typically reported for methods benchmarking are based on AUC because of the reasons listed above. Thus, if many benchmarks use ROC-based quality metrics and one wants to surpass current methods in these benchmarks,  one can ideally use cost functions that directly aim to optimize those quality metrics.

\paragraph{PU learning}
The machine learning task where only positive and unlabeled samples are available for training is referred to as \emph{positive/unlabeled learning} (PU learning).
An example application of PU learning in drug discovery is the classification of small molecules into drug-like and non-drug-like compounds. While we can assemble lists of drugs and known drug-like compounds as positive training samples, it is difficult to come up with lists of definitely non-drug-like compounds -- or at least not any that are not trivially non-drug-like. However, we can come up with sets of unlabeled compounds with unknown drug likeness.
\citet{zhang2008learning} show that PU learning is equivalent to treating the unlabeled samples as negative while optimizing an AUC-based cost function.
\citet{Ren2018PU} show the effectiveness of AUC maximization for highly imbalanced PU learning for the special case where a part of the training data is mislabeled and/or some features are redundant.

\subsection{Difficulties of AUC optimization and related work}
While there are important reasons to optimize the AUC, it also gives rise to some challenging problems.
Firstly, the gradient of AUC with respect to network weights is zero almost everywhere, because AUC changes only when the ranks of predictions for a positive sample and a negative sample are swapped. Therefore, AUC cannot be optimized directly by first-order optimization methods.
Secondly, the AUC is a sum over pairs of samples rather than over individual samples, making its direct computation slow. In the following we outline typical solutions to these problems.

\paragraph{Approximations to AUC with non-zero gradient}
Different approximations to the AUC have been proposed that have non-zero gradients (on more than only a null set of weight space), allowing gradient-based optimization.

A sum of sigmoidal functions with the arguments scaled by a pre-defined value is a good approximation to the AUC \citep{yan2003optimizing}. However, it comes at the cost of creating very steep gradients depending on the choice of this scaling value, making optimization difficult. One of the differentiable AUC approximations introduced by \citet{yan2003optimizing} does not have the issue of steep gradients. The proposed objective function dynamically adjusts a sample pairs' contribution to the loss, based on the score difference between the positive and the negative sample. More specifically, if the difference is larger than a specified margin, then the contribution of a given pair to the loss is zero, otherwise it is a positive value that changes smoothly with the magnitude of the score difference. This formulation makes the optimization focus on maximizing the number of pairs that have a pairwise difference larger than a given margin, enhancing generalization performance. For these reasons, we base our proposed objectives on this loss function (named $R_1$) by \citet{yan2003optimizing}.

\textit{RankOpt} \citep{herschtal2004optimising} is a linear classifier which uses sigmoidal functions in the cost function with the scaling value being calculated from the data instead of fixing it a priori.
The algorithm is also computationally efficient, making it linear in the number of samples as opposed to quadratic run time of other methods.
However, being a linear classifier, RankOpt is not directly compatible with deep learning.

The AUC ignores the exact prediction scores (which might contain valuable additional information about the model's quality, for example scoring a positive as ``$0.9$ (rank~$1$)'' might indicate a more promising model than scoring it as ``$0.7$ (rank~$1$)''). The AUC takes only ranks into account. (This limitation of the AUC also causes the zero-gradient problem.)
To overcome this issue, a method called \textit{scored ROC} was proposed \citep{Wu2007} which is based on reducing scores for positives by a number between $0$ and $1$, the so-called margin. To this end, the \emph{scored ROC curve} (not similar to ROC) plots margins against the corresponding AUC. The area under the scored ROC curve, called \emph{scored AUC}, measures how quickly AUC declines when classifier outputs for positives are reduced. This metric has non-zero gradients with respect to sample scores.

\citet{calders2007efficient} use a polynomial (rather than sigmoidal) approximation of the step function. Their approximation allows to reformulate the sum over pairs of samples into a sum over individual samples, thus reducing the runtime of an epoch from quadratic to linear complexity. We achieve such runtime reduction with a sigmoidal approximation by using a lookup table that maps decision thresholds to false positive rates (see Section~\ref{sec:lookup}).
\citet{Ferri2005} use a linear (rather than sigmoidal or polynomial) approximation of the step function.
A study of the bias of several AUC approximations was performed by \citet{Vanderlooy2016}.

\paragraph{Direct optimization of AUC without gradient-based methods}
Optimizing AUC directly using coordinate descent (i.e.\ optimizing one model parameter at a time, which is feasible despite the zero gradient) yields good results for certain machine learning methods that were designed specifically for genetics applications \citep{zhu2017direct}.
\citet{LeDell2016AUCmax} introduce an ensemble approach based on \textit{Super Learner} \citep{vanderLaan2007} which adjusts the combination of scores from several individual classifiers in favor of a higher AUC.
They show that even though none of the base classifiers is specifically trained to maximize AUC, the Super Learner ensemble outperforms the top base algorithm, especially on data with high class imbalance.

\paragraph{Online AUC optimization}
We use out-of-batch predictions (see Section~\ref{sec:aucprev}) and a lookup table (see Section~\ref{sec:lookup}). These techniques are related to online AUC optimization (i.e.~training where data are not available all at once),
which rewrites the sum of losses over sample pairs into a sum of losses of individual samples and uses buffers for positive and negative training samples \citep{zhao2011online} or stores the first- and second-order statistics of training data \citep{gao2013one}.

\section{Methods}
\label{sec:methods}
In this section we will first introduce the AUC-based cost functions upon which our work relies, then we introduce the novel cost functions. The main contributions are: 
  identification of five reasons for using ROC-based cost functions in virtual screening (Section~\ref{sec:reasons}),
  a novel algorithm \textit{AUC-prev}
  that uses out-of-batch predictions for overall prediction improvement in AUC optimization (Section~\ref{sec:aucprev}),
  and
  a new cost function $\LlogAUC$ that optimizes the ROC curve using a novel reweighting scheme for different decision thresholds, along with a lookup table for fast computation and a stop-gradient operator that prevents degenerate solutions (Section~\ref{sec:logauc}), and a comparison of five cost functions using four quality metrics and nine representative drug discovery datasets.

\subsection{Approximation of the AUC}

The AUC is the ratio of all positive-negative pairs where the positive has a higher prediction than the negative. In other words, 
\begin{equation}\label{eq:AUC}
\textnormal{AUC} = \frac{\sum_{i=1}^m \sum_{j=1}^n H(x_i-y_j)}{m n},
\end{equation}
where $H(z) = \mathbbm{1}[z > 0]$ is the Heaviside step function, the $x_i$ are all $m$ positive samples and the $y_j$ are all $n$ negative samples.

\citet{yan2003optimizing} proposed an approximation to $H(x_i-y_j)$ with partially non-zero gradients:
\begin{equation}\label{eq:f}
f(x_i, y_j) = \ifelse{(-(x_i - y_j - \gamma))^p}{x_i - y_j < \gamma}{0},
\end{equation}
where $0 < \gamma \le 1$ (usually $0.1 \le \gamma \le 0.7$) and $p > 1$ (usually $2$ or $3$) are hyperparameters.
When comparing two classifiers that have the same AUC value, one of them may be better in the sense that it separates the positive scores from the negative scores by a larger margin. Equation~\eqref{eq:f} incorporates the margin $\gamma$ to distinguish between these cases. The average of $f(x_i, y_j)$ taken over all pairs $(x_i, y_j)$, i.e.
\begin{equation}\label{eq:uAUC}
\LAUC = \frac{1}{mn} \sum_{i=1}^m \sum_{j=1}^n f(x_i, y_j),
\end{equation}
is a cost function that encourages positive samples to have a score that is higher by at least $\gamma$ than the score for negative samples. 
If this condition is not met, then that particular positive-negative pair contributes to the loss.

\subsubsection{Lower-left part ROC curve optimization}

The \emph{left} part of the ROC changes if positive-negative pairs that have relatively \emph{high} scores swap ranks.
This portion of the ROC curve describes the classifier performance at high decision thresholds, i.e.~when selecting only the top few percent of candidates, which is usually the case in drug discovery due to the high cost of subsequent \emph{in vitro} experiments.
To optimize for such situations, \citet{yan2003optimizing} further modify Eq.~\eqref{eq:uAUC} to transform the scores of positive and negative samples by the function $g(\cdot)$ before passing the pair to the cost function:
\begin{equation}\label{eq:uleftAUC}
\LleftAUC = \frac{1}{mn} \sum_{i=1}^m \sum_{j=1}^n f(g(x_i), g(y_j)),
\end{equation}
\begin{equation}\label{eq:g}
\text{where } g(s) = \ifelse{( s - \beta \mu_s )^\alpha}{s > \beta \mu_s}{0},
\end{equation}
with hyperparameters $\alpha > 1$ but close to $1$ (usually $1.1$) and $\beta \ge 1$ (usually $1$), and
$\mu_s$ is the mean value of the classifier scores for all samples. Positive samples that have a high score will be mapped to a value that depends on the magnitude of the difference between the score of that sample and the classifier mean score.
This has an effect of pushing these samples even more in the direction of high positive classification.

\subsection{New cost functions}

Based on the AUC and its approximation, Eq.~\eqref{eq:uAUC}, we propose two new objective functions
for training any parametric classifiers to directly optimize the ROC curve using gradient-based methods. We use them with a multilayer perceptron with softmax outputs for applications in drug discovery.

\subsubsection{AUC optimization using out-of-batch predictions}
\label{sec:aucprev}
A mini-batch that consists of $0.1\%$ of all positive and negative\footnote{The datasets we use have very few negative samples (not only in terms of relative class imbalance, but also in terms of absolute numbers), i.e.~the advantages of the following method can be expected to be even more pronounced on larger (including imbalanced) datasets.} samples computes 
$0.1\%$ of the overall cost for usual cost functions, but only $0.0001\%$ (if pairs are coherent, see Section~\ref{sec:coherent}) or less for cost functions that are a sum over pairs of samples such as ROC-based ones. Thus, one ``epoch'' (i.e.~seeing the overall loss once) requires an enormous number of mini-batches (quadratic rather than linear in the number of samples), and quick convergence to a good solution can be problematic. To address this problem, we propose using out-of-batch predictions for overall prediction improvement. In each iteration, the newest predictions for samples from the current mini-batch are used to update the weights of the network accordingly, while the \emph{recent} predictions for all samples (including out-of-batch ones) are also used for loss computation, but considered constants that do not depend on the network weights. The proposed mini-batch-wise objective function looks as follows:
\begin{equation}\label{eq:uAUC-prev}
\LAUCprev = \LAUC(\Xcurr, \Ycurr) + \LAUC(\Xcurr, \Yprev)  + \LAUC(\Xprev, \Ycurr) ,
\end{equation}
where $\Xcurr$, $\Ycurr$ are current predictions for positive resp.\ negative samples from the current mini-batch 
and $\Xprev$, $\Yprev$ are the most recent predictions for \emph{all} positive resp.\ negative samples. 
With this cost function,
the network weights are optimized by not only considering the loss contribution of positive-negative pairs from the current mini-batch but also from pairs which consist of a current-mini-batch sample and an out-of-batch sample. Compared to other approaches, this procedure allows the loss to be based upon more samples than present in the mini-batch, reducing noise in the loss and making the training more stable.

\subsubsection{Optimizing the area under the lin-log ROC curve}

\label{sec:logauc}
If the goal is to optimize for early enrichment, i.e.~not missing out on good candidates at high decision thresholds, then a suitable performance measure is the area under a part of the lin-log ROC curve, called \emph{logAUC} \citep{mysinger2010rapid}. This quality measure shares some properties with the AUC statistic (e.g. robustness to class imbalance) but is biased towards early enrichment.
The logAUC metric is often used for measuring the quality of methods that were trained with other metrics such as cross-entropy.
Defining quality differently during training than during evaluation is suboptimal.
We develop a new cost function targeted at directly optimizing the logAUC metric.

We observe that the AUC, Eq.~\eqref{eq:AUC}, can be interpreted as integration over the ROC curve, which consists of $n$ stripes of equal width:
\begin{align}
\textnormal{AUC} 
&= \frac{1}{mn} \sum_{j=1}^n \sum_{i=1}^m H(x_i-y_j) \\
&= \sum_{j=1}^n\left[\frac{j+1}{n}-\frac{j}{n}\right] \frac{1}{m} \sum_{i=1}^m H(x_i-y_j), \label{eq:AUCstripes}
\end{align}
where we assume the negatives to be sorted in order of descending classifier output. The term $\frac{1}{m} \sum_{i=1}^m H(x_i-y_j)$ corresponds to the TPR at the decision threshold $y_j$, i.e.\ to the height of the $j^\textnormal{th}$ stripe, and $\left[\frac{j+1}{n}-\frac{j}{n}\right] = \frac{1}{n}$ is the stripe width ($\frac{j}{n}$ and $\frac{j+1}{n}$ are the left and right coordinates of the stripe, respectively). So $\left[\frac{j+1}{n}-\frac{j}{n}\right] \frac{1}{m} \sum_{i=1}^m H(x_i-y_j)$ is the area of the $j^\textnormal{th}$ stripe, and the outer sum goes over all $n$ stripes.

Computing the area under the part of the curve where FPR is in $[\lambda; 1]$ (with e.g.~$\lambda=0.001$) instead of $[0;1]$ can be done as follows:
\begin{equation}
    \t{AUC}_\lambda = \sum_{j=1}^n\left[
        \t{clip}_{[\lambda;1]}\left(\frac{j+1}{n}\right) -
        \t{clip}_{[\lambda;1]}\left(\frac{j}{n}\right)
    \right] \frac{1}{m} \sum_{i=1}^m H(x_i-y_j),
\end{equation}
where $\t{clip}_{[a;b]}(z)=\min\{\max\{z,a\},b\}$ clips the abscissa coordinates to the interval $[\lambda;1]$.

The area logAUC under the log-transformed ROC curve can be computed by transforming the abscissa coordinates (in square brackets) to logarithmic scale:
\begin{equation}\label{eq:logAUC}
    \textnormal{logAUC}_\lambda = \sum_{j=1}^n
    \left[\log \left( \t{clip}_{[\lambda;1]} \left( \frac{j+1}{n}\right)\right)-\log \left( \t{clip}_{[\lambda;1]} \left( \frac{j}{n}\right)\right)\right]
    \frac{1}{m} \sum_{i=1}^m H(x_i-y_j).
\end{equation}
Clipping to $[\lambda;1]$ is essential because otherwise logAUC would be infinite.

Equivalently to the explanation for AUC above, logAUC has gradient zero almost everywhere. In order to optimize logAUC with gradient-based methods, we approximate it by replacing the step function $H$ in Eq.~\eqref{eq:logAUC} by a smooth  function, as was done for AUC in Eqs.~\mbox{\eqref{eq:AUC}--\eqref{eq:uAUC}}. Defining $w_j := \log \left( \t{clip}_{[\lambda;1]} \left( \frac{j+1}{n}\right)\right)-\log \left(\t{clip}_{[\lambda;1]} \left( \frac{j}{n}\right)\right)$, we obtain the logAUC objective function
\begin{equation}\label{eq:LlogAUC}
    \LlogAUC = \sum_{j=1}^n \left( w_j \ \sum_{i=1}^m \ f(x_i, y_j) \right) = \sum_{j=1}^n \sum_{i=1}^m w_j \ f(x_i, y_j),
\end{equation}
where the weighting factor $w_j$ from Eq.~\eqref{eq:logAUC} was pulled into the inner sum such that each positive-negative pair has a weighting factor for batch-based training.

This cost function directly optimizes the logAUC metric by individually scaling each of the $n$ equal-width stripes which the AUC is composed of (like the logAUC metric does by log-transforming the abscissa of the ROC), thus giving more importance to the left part of the ROC curve.

\paragraph{Acceleration of rank computation}
\label{sec:lookup}
There are a couple of implementation details worth mentioning.
First, the usage of the weighting factor $w_j$ requires computing the rank $j$ of each negative sample (see the assumption under Eq.~\eqref{eq:AUCstripes}) from its prediction $y_j$, which in turn requires sorting all negatives by their predictions. To save time, we construct a lookup table which maps thresholds to FPRs, allowing us to then estimate a negative's rank $j$.
The lookup table is updated after every epoch (pass through all pairs) and uses equidistant keypoints in threshold space. Lookup uses linear interpolation. We found a step size of $0.001$ to be a good trade-off between accuracy and time requirements.

This imprecise rank computation leads to imprecise cost gradients and degenerate solutions. Details and a remedy are described in the following.
The score of each sample in a mini-batch is between two adjacent entries in the lookup table and we interpolate linearly between these two thresholds. Smaller predictions for \emph{negatives} lead (by changing the interpolation weights)
to higher estimates of sample ranks (because the prediction-to-rank mapping is a monotonically decreasing mapping). This causes
the estimates of the sample weights $w_j$ (estimated stripe widths in log space) to decrease. Decreasing estimates of $w_j$ lead to a decreasing estimate of the overall loss, cf.~Eq.~\eqref{eq:LlogAUC}. As a smaller loss is preferred by the optimization algorithm, and gradients can flow through this entire pipeline
, predicted scores for negatives are incentivized to become smaller and converge to zero very quickly. Also the scores for positives converge to zero, apparently as a side effect, because the network does not have sufficient time/incentive to learn to distinguish them and simply learns to always output zero.
The remedy is during an update step to
conceal the strictly monotonous dependence of \emph{rank estimates} on predictions, mimicking the zero gradient of the piecewise constant dependence of \emph{actual ranks} on predictions.
To this end, we use the stop-gradient operator $\t{sg}[\cdot]$ which sets gradients to zero when back-propagating through it. We replace $w_j$ by $\t{sg}[w_j]$. This prevents degenerate solutions.

\subsection{Experimental setup}
\paragraph{Datasets}
\label{sec:dataset}
Results are reported on nine large Quantitative Structure--Activity Relationship (QSAR) benchmark datasets \citep{butkiewicz2013benchmarking} comprising small molecules labeled as active or inactive for nine protein targets.
Each dataset was derived from a single screening effort in the PubChem database, and label accuracy for actives was confirmed via validating screens. As a cohesive set, these datasets avoid the construction biases often seen in other datasets, which can result in trivial decision boundaries and poor generalization (something which also affects \citep{chen2019hidden} typical docking datasets such as DUD-E). They also display the realistically large numbers of diverse compounds and the class imbalances (a few hundred active molecules and inactives on the order of $10^5$) typically seen in practical drug development projects, properties often lacking in model systems. See~\citet{butkiewicz2013benchmarking} for details.
As input features, we used the 391 descriptors from the Reduced Short Range descriptor set, which has previously been identified as sufficiently informative and compact~\citep{mendenhall2016improving,vu2019bcl}.
We normalized features using z-score scaling,
which was found to be the most effective for these datasets~\citep{mendenhall2016improving}.

\paragraph{Network architecture}
For our experiments we adopt many of the architecture design choices that have proven to be useful~\citep{mendenhall2016improving} on the same PubChem QSAR datasets with standard cost functions. Throughout the experiments we use a two-layer feed-forward neural network with one hidden layer of thirty-two neurons with rectified linear units as activation functions and one output unit with sigmoidal activation function. Also, we use dropout in the input layer and in the first hidden layer. We use dropout rates that were optimized by \citet{mendenhall2016improving} for each of the nine datasets individually.
Using dropout has been proven to be much more beneficial for QSAR datasets than employing additional hidden layers or larger neural networks, whose effect is shown to be insignificant in comparison~\citep{mendenhall2016improving}.

\paragraph{Training procedure}
\label{sec:coherent}
We train on mini-batches which are constructed as follows: each mini-batch contains coherent positive-negative pairs, i.e.~we randomly uniformly draw positive and negative samples from the training set and construct all possible pairs between these. The use of coherent mini-batches has many advantages such as being memory-efficient, easy to implement, and parallelizable.

All parameters are initialized using the scheme by \citet{he2015delving}. For training, we used the Adam optimization algorithm~\citep{kingma2014adam} with different learning rates for each task, exponential decay rate for the first and second moment at $0.9$ and $0.999$, respectively. The optimal learning rates were found by $K$-fold cross-validation and testing approach as in~\citep{korjus2016efficient} with $K=4$. The best learning rate was $0.001$ for all nine tasks and all objective functions, except for $\LlogAUC$ on dataset ID 2258, where the best learning rate was $0.003$.

\section{Results and discussion}

Similar to~\citep{mysinger2010rapid}, model performance is evaluated by computing logAUC for FPR in $[0.001; 0.1]$. This quality metric is very popular for chemistry-related problems such as drug discovery, where we focus on high decision thresholds.
In addition to this quality metric we report two other metrics that focus on high decision thresholds,
namely AUC for FPR in $[0.001; 0.1]$ and logAUC for FPR in $[0.001; 1]$; as well as AUC.

\begin{table}[t]
 \caption{Evaluation using ``AUC'' as a quality metric. The loss functions $\LAUC$ and $\LAUCprev$ proposed for optimizing this quality metric are highlighted in black, other loss functions in \textcolor{otherloss}{grey}. Results that significantly ($p<0.05$) outperform the baseline method $\Lce$ are marked \textbf{bold}. On all 4 datasets on which $\LAUCprev$ outperforms $\Lce$, it also outperforms $\LAUC$, indicating advantages of our proposed usage of out-of-batch samples over usual batch-wise training.
 Error margins range from $\pm 0.0004$ to $\pm 0.0019$. }
 \label{table:AUC}
 \centering
 \begin{tabular}{l|gllgg}
\toprule
SAID & 
\rotatebox[origin=lB]{90}{$\Lce$} & 
\rotatebox[origin=lB]{90}{$\LAUC$} & 
\rotatebox[origin=lB]{90}{$\LAUCprev$} & 
\rotatebox[origin=lB]{90}{$\LleftAUC$} & 
\rotatebox[origin=lB]{90}{$\LlogAUC$} 
\\
\midrule                      
1798   &     0.727 & \bf 0.767 & \bf 0.769 & \bf 0.759 & \bf 0.752 \\
1834   &     0.939 & \bf 0.944 &     0.934 &     0.936 & \bf 0.951 \\
2258   &     0.809 & \bf 0.819 & \bf 0.826 &     0.810 &     0.773 \\
2689   &     0.870 & \bf 0.877 & \bf 0.878 &     0.859 &     0.865 \\
435008 &     0.785 &     0.778 &     0.770 &     0.772 &     0.775 \\
435034 &     0.814 & \bf 0.816 &     0.808 & \bf 0.818 & \bf 0.832 \\
463087 &     0.861 &     0.860 &     0.859 &     0.846 &     0.862 \\
485290 &     0.753 &     0.756 &     0.748 &     0.748 & \bf 0.760 \\
488997 &     0.767 &     0.769 & \bf 0.773 &     0.762 &     0.763 \\
\bottomrule
\end{tabular}
\end{table}

\begin{table}[t]
 \caption{Evaluation using ``AUC for FPR in $[0.001; 0.1]$'' as a quality metric. The loss function $\LleftAUC$ designed to optimize this quality metric is highlighted in black. It significantly ($p<0.05$, \textbf{bold}) outperforms $\Lce$ on 4 out of 9 datasets. $\LlogAUC$, designed for a similar purpose, significantly outperforms $\Lce$ on 5 out of 9 datasets.
 Error margins range from $\pm 0.0011$ to $\pm 0.0028$. }
 \label{table:pAUC}
 \centering
 \begin{tabular}{l|ggglg}
\toprule
SAID &      
\rotatebox[origin=lB]{90}{$\Lce$} & 
\rotatebox[origin=lB]{90}{$\LAUC$} & 
\rotatebox[origin=lB]{90}{$\LAUCprev$} & 
\rotatebox[origin=lB]{90}{$\LleftAUC$} & 
\rotatebox[origin=lB]{90}{$\LlogAUC$} 
\\
\midrule                      
1798   &     0.272 & \bf 0.315 &     0.260 &     0.247 &     0.211 \\
1834   &     0.622 & \bf 0.653 &     0.606 &     0.591 & \bf 0.662 \\
2258   &     0.523 & \bf 0.547 &     0.489 &     0.511 &     0.298 \\
2689   &     0.589 & \bf 0.621 & \bf 0.603 & \bf 0.601 &     0.589 \\
435008 &     0.370 & \bf 0.376 &     0.361 & \bf 0.389 &     0.364 \\
435034 &     0.388 &     0.388 &     0.386 &     0.380 & \bf 0.401 \\
463087 &     0.342 &     0.338 &     0.339 &     0.318 & \bf 0.362 \\
485290 &     0.392 &     0.393 &     0.385 & \bf 0.403 & \bf 0.418 \\
488997 &     0.383 & \bf 0.451 & \bf 0.444 & \bf 0.453 & \bf 0.432 \\
\bottomrule
\end{tabular}
\end{table}

Tables~\ref{table:AUC}--\ref{table:plogAUC} report these four quality metrics for each of the four ROC-based objective functions. Additionally, results are compared to the baseline method, i.e.~the cross-entropy objective. Each row represents one of the nine datasets. Bold faced numbers indicate that the corresponding objective function
significantly ($p<0.05$)
outperforms the baseline for that particular dataset. 
Confidence intervals for each metric were computed by bootstrapping the test set with replacement 200 times \citep{mendenhall2016improving}. For each metric, the results of
the cost functions proposed specifically for that metric are shown in black. The results of other cost functions are also shown, but greyed out.

Table~\ref{table:AUC} shows results using the AUC quality metric. The cost functions designed specifically to optimize this metric are $\LAUC$ and $\LAUCprev$. The cost function $\LAUC$ outperforms the baseline method $\Lce$ at a significance level of $\alpha=0.05$ on 5 out of the 9 datasets. Our procedure $\LAUCprev$ of using out-of-batch predictions outperforms the baseline method significantly ($\alpha=0.05$) on 4 out of 9 datasets, including one dataset on which $\LAUC$ does not outperform $\Lce$.
Moreover, on all 4 datasets on which $\LAUCprev$ outperforms $\Lce$, it also outperforms $\LAUC$ (with $p<0.05$ on two of the datasets),
indicating advantages of our proposed usage of out-of-batch samples.

Results in Table~\ref{table:pAUC} are evaluated using the ``AUC for FPR in $[0.001; 0.1]$'' quality metric. The function $\LleftAUC$ was designed specifically for optimizing this metric. The results show that $\LleftAUC$ outperforms the cross-entropy baseline in 4 out of 9 datasets at a significance level of $\alpha=0.05$. Our objective function $\LlogAUC$ (designed for a similar purpose) significantly outperforms $\Lce$ in 5 out of 9 datasets. This indicates that an ROC-based cost function that maximizes the area under the left part of the ROC can improve classifier performance as compared to typical cost functions (such as cross-entropy) when the goal is to optimize the performance at high decision thresholds.
In addition, $\LlogAUC$ significantly outperforms $\LleftAUC$ in 4 out of 9 datasets.

\begin{table}[t]
 \caption{Evaluation using ``logAUC for FPR in $[0.001; 1]$'' as a quality metric. The loss function $\LlogAUC$ designed to optimize this quality metric significantly ($p<0.05$, \textbf{bold})  outperforms $\Lce$ on 7 out of 9 datasets.
 Error margins range from $\pm 0.0007$ to $\pm 0.0022$.}
 \label{table:logAUC}
 \centering
 \begin{tabular}{l|ggggl}
\toprule
SAID & 
\rotatebox[origin=lB]{90}{$\Lce$} & 
\rotatebox[origin=lB]{90}{$\LAUC$} & 
\rotatebox[origin=lB]{90}{$\LAUCprev$} & 
\rotatebox[origin=lB]{90}{$\LleftAUC$} & 
\rotatebox[origin=lB]{90}{$\LlogAUC$}
\\
\midrule                                                                           
1798   &     0.329 & \bf 0.353 & \bf 0.333 &     0.324 &     0.298 \\
1834   &     0.575 & \bf 0.589 &     0.557 &     0.562 & \bf 0.629 \\
2258   &     0.489 & \bf 0.514 &     0.476 &     0.482 &     0.411 \\
2689   &     0.531 & \bf 0.545 & \bf 0.539 & \bf 0.538 & \bf 0.558 \\
435008 &     0.391 & \bf 0.396 &     0.390 & \bf 0.403 & \bf 0.399 \\
435034 &     0.415 &     0.415 &     0.411 &     0.415 & \bf 0.430 \\
463087 &     0.397 &     0.398 &     0.393 &     0.385 & \bf 0.404 \\
485290 &     0.418 & \bf 0.425 &     0.411 &     0.419 & \bf 0.427 \\
488997 &     0.395 & \bf 0.425 & \bf 0.417 & \bf 0.425 & \bf 0.440 \\
\bottomrule
\end{tabular}
\end{table}

\begin{table}[t]
 \caption{Evaluation using ``logAUC for FPR in $[0.001; 0.1]$'' as a quality metric. The loss function $\LlogAUC$ designed to optimize this quality metric significantly ($p<0.05$, \textbf{bold}) outperforms $\Lce$ on 7 out of 9 datasets.
 Error margins range from $\pm 0.0008$ to $\pm 0.0025$.}
 \label{table:plogAUC}
 \centering
 \begin{tabular}{l|ggggl}
\toprule
SAID & 
\rotatebox[origin=lB]{90}{$\Lce$} & 
\rotatebox[origin=lB]{90}{$\LAUC$} & 
\rotatebox[origin=lB]{90}{$\LAUCprev$} & 
\rotatebox[origin=lB]{90}{$\LleftAUC$} & 
\rotatebox[origin=lB]{90}{$\LlogAUC$}
\\
\midrule                                                                           
1798   &     0.154 & \bf 0.164 &     0.143 &     0.141 &     0.101 \\
1834   &     0.390 & \bf 0.408 &     0.367 &     0.372 & \bf 0.466 \\
2258   &     0.339 & \bf 0.372 &     0.306 &     0.330 &     0.238 \\
2689   &     0.367 & \bf 0.382 & \bf 0.373 & \bf 0.384 & \bf 0.413 \\
435008 &     0.219 & \bf 0.229 & \bf 0.224 & \bf 0.241 & \bf 0.231 \\
435034 &     0.229 &     0.227 &     0.228 &     0.226 & \bf 0.245 \\
463087 &     0.176 &     0.175 &     0.171 &     0.171 & \bf 0.189 \\
485290 &     0.268 & \bf 0.279 &     0.265 & \bf 0.277 & \bf 0.281 \\
488997 &     0.228 & \bf 0.268 & \bf 0.259 & \bf 0.275 & \bf 0.297 \\
\bottomrule
\end{tabular}
\end{table}

Table~\ref{table:logAUC} demonstrates results for the ``logAUC for FPR in $[0.001; 1]$'' quality metric. Our proposed objective function $\LlogAUC$ was specifically designed for this metric and performs significantly better ($\alpha=0.05$) than cross-entropy on 7 out of 9 datasets. The results also show that $\LlogAUC$ outperforms other ROC-based objective functions on many of the datasets. Specifically, it significantly outperforms $\LAUC$ on 5 out of 9 datasets.

Table~\ref{table:plogAUC} shows results for the ``logAUC for FPR in $[0.001; 0.1]$'' quality metric which is also optimized by our $\LlogAUC$ objective. The results demonstrate that, again, our objective function outperforms the cross-entropy baseline in 7 out of 9 datasets and $\LAUC$ in 5 out of 9 datasets, both at a significance level of $\alpha=0.05$.

One important aspect to mention is that when using $\Lce$ we oversample the minority class (positives) in order to compare the AUC-based cost functions against the cross-entropy under ideal circumstances. Thus we should expect even more favorable results compared with cross-entropy if no oversampling is performed.
The AUC-based cost functions are robust towards class imbalance, i.e.~perform equally well without oversampling.
Thus, they do not require tuning the oversampling ratio, as the cross-entropy loss does. On the other hand, they have additional hyperparameters that require tuning. Luckily, a wide range of values works well in practice~\citep{yan2003optimizing}.

\section{Conclusions}
\label{sec:conclusionfuturework}

We listed a series of special properties of virtual screening datasets, such as class imbalance, all of which can be addressed by using ROC-based cost functions. Such cost functions, in turn, have peculiarities such as the necessity for techniques to avoid zero gradients (which we borrowed from literature), and a quadratic rather than linear number of summands (which we addressed by proposing to use out-of-batch samples and coherent mini-batches). Moreover, to optimize performance specifically for the high decision thresholds that are used in virtual screening, and to more directly optimize the logAUC quality metric that is popular in this domain, we proposed an approximation $\LlogAUC$ to logAUC with nonzero gradients. To accelerate its computation, we replaced precise computation by a lookup table, and to prevent the wrong gradient caused by this replacement from leading to degenerate solutions, we used a stop-gradient operator.
Our methods outperformed cross-entropy in many scenarios in a benchmark of realistic diverse datasets.
We do not claim that these AUC losses are perfect for all situations; rather, we exemplify how losses can be aligned with project-specific goals. We encourage active exploration of loss options in virtual screening and in other applications, instead of following the old tradition of resorting to cross-entropy ``by default''.

\section*{Acknowledgements}
We thank Benjamin~P.~Brown for helpful discussions.

\bibliographystyle{abbrvnat}
\begin{small}
\bibliography{refs}
\end{small}

\end{document}